\documentclass[runningheads]{llncs} 
\pdfoutput=1
\usepackage[utf8]{inputenc}
\usepackage{algorithm} 
\usepackage{algpseudocode}  
\usepackage[english]{babel}
\usepackage{color}
\usepackage{colortbl,tabulary,tabularx,multirow}
\usepackage{ltxtable}
\usepackage{comment,paralist}
\usepackage{wrapfig}
\usepackage{amsmath}
\usepackage{amssymb}
\usepackage{stmaryrd}
\usepackage{braket}
\usepackage{textcomp}
\usepackage{listings}
\usepackage{theorem}
\usepackage{alltt}
\usepackage{epsfig,psfrag,subfigure}
\usepackage{graphicx}
\usepackage[colorlinks,linkcolor=blue,citecolor=blue,urlcolor=blue]{hyperref}
\usepackage{multicols}
\usepackage{balance}
\usepackage{amssymb}

\newcommand{\Z}{\mathbb{Z}}
\newcommand{\R}{\mathbb{R}}

\everymath{\displaystyle}

\usepackage{tikz}
\usetikzlibrary{shapes,arrows,decorations.pathmorphing,backgrounds,fit,matrix,calc}

\definecolor{todo-color}{rgb}{1,0,0}
\newcommand{\todo}[1]{\textcolor{todo-color}{[#1]}}

\definecolor{ploc-color}{rgb}{0.9,0.6,0}

\definecolor{romain-color}{rgb}{1,0,1}

\definecolor{pierre-color}{rgb}{0,1,1}

\definecolor{no-color}{rgb}{1,0.2,0.2}

\renewcommand{\c}{\ensuremath{\operatorname{c}\!}}

\newcommand{\benum}{\begin{enumerate}}
\newcommand{\eenum}{\end{enumerate}}
\newcommand{\nc}{\newcommand}
\newcommand{\rnc}{\renewcommand}
\newcommand{\bvb}{\begin{verbatim}}
\newcommand{\evb}{\end{verbatim}}

\newcommand{\Lb}{\llbracket} 
\newcommand{\Rb}{\rrbracket} 
\newcommand{\Ra}{\Rightarrow}

\nc{\bq}{\textbf}
\nc{\m}{\textrm}
\nc{\bb}{\mathbb}
\nc{\til}{\texttildelow}
\nc{\be}{\begin{equation}}
\nc{\ee}{\end{equation}}
\nc{\dps}{\displaystyle}
\rnc{\l}{\left(}\rnc{\r}{\right)}
\nc{\lc}{\left\{}\nc{\rc}{\right\}}
\nc{\lb}{\left[}\nc{\rb}{\right]}
\nc{\ba}[1]{\begin{array}{#1}}
\nc{\ea}{\end{array}}       
\nc{\ra}{\rightarrow}
\nc{\li}{\left |}
\nc{\ri}{\right |}
\nc{\pde}[2]{\frac{\partial #1}{\partial #2}}
\nc{\ode}[2]{\frac{d #1}{d #2}}
\nc{\odee}[3]{\frac{d^{#3} #1}{d #2^{#3}}}
\nc{\pdee}[3]{\frac{\partial^{#3} #1}{\partial #2^{#3}}}
\nc{\bn}{\begin{enumerate}}
\nc{\en}{\end{enumerate}}
\nc{\bt}{\begin{theorem}}
\nc{\et}{\end{theorem}}
\nc{\y}[1]{\lambda_{#1}}
\nc{\ninf}{{\oplus}^{-\infty}}
\nc{\pinf}{{\oplus}^{+\infty}}
\nc{\nninf}{{\otimes}^{-\infty}}
\nc{\ppinf}{{\otimes}^{+\infty}}
\nc{\ir}{\mathbb{I}\mathbb{R}}

\nc{\ep}{\mathcal{E}_{P}}
\nc{\mr}{\mathcal{M}_{r}}
\nc{\mfa}{\mathcal{M}_{f,a}}
\nc{\mfp}{\mathcal{M}_{f,p}}
\nc{\mt}{\m{T}}
\nc{\F}{\mathbb{F}}

\tikzstyle{block} = [draw,rectangle,thick,minimum height=2em,minimum width=\textwidth]
\tikzstyle{sum} = [draw,circle,inner sep=0mm,minimum size=2mm]
\tikzstyle{connector} = [->,thick]
\tikzstyle{line} = [thick]
\tikzstyle{branch} = [circle,inner sep=0pt,minimum size=1mm,fill=black,draw=black]
\tikzstyle{guide} = []
\tikzstyle{snakeline} = [connector, decorate, decoration={pre length=0.2cm,
                         post length=0.2cm, snake, amplitude=.4mm,
                         segment length=2mm},thick, magenta, ->]

\usepackage{environ}

\def\comment#1{}
\newlength{\hsbw}

\tikzstyle{mybox} = [draw, very thick, rectangle, rounded corners, inner sep=0pt, inner ysep=2pt]
\tikzstyle{fancytitle} =[fill=white, draw, rectangle, rounded corners, very thick]
\newsavebox{\GrayRoundedBox}
\NewEnviron{mytikzbox}[1]{%
  \begin{lrbox}{\GrayRoundedBox}
    \setlength{\hsbw}{\linewidth}
    \addtolength{\hsbw}{-8pt}
    \begin{minipage}[b]{\hsbw}
      \begingroup\small
      \begin{alltt}
        \BODY
      \end{alltt}
      \endgroup
    \end{minipage}
  \end{lrbox}
  \begin{flushleft}
    \vspace{-10pt}
    \begin{tikzpicture}
      \node[mybox](box){\usebox{\GrayRoundedBox}};
      \node[fancytitle, left=10pt] at (box.north east) {\tiny {#1}};
    \end{tikzpicture}
  \end{flushleft}
}

\newsavebox{\Zname}
\newenvironment{tikzboxtt}[1]{%
  \sbox\Zname{\tiny {#1}}
  \begin{lrbox}{\GrayRoundedBox}
    \setlength{\hsbw}{\linewidth}
    \addtolength{\hsbw}{-8pt}
    \begin{minipage}[b]{\hsbw}
      \begingroup\small
      \begin{alltt}
}{    \end{alltt}
      \endgroup
    \end{minipage}
  \end{lrbox}
  \begin{flushleft}
    \vspace{-1em}
    \begin{tikzpicture}
      \node[mybox](box){\usebox{\GrayRoundedBox}};
      \node[fancytitle, left=10pt] at (box.north east) {\usebox{\Zname}};
    \end{tikzpicture}
  \end{flushleft}
  \vspace{-.5em}}

\tikzset{diagram background/.style={fill=blue!5,rounded corners=0.5cm}}
\tikzstyle{block} = [rectangle, draw, fill=blue!20, 
    minimum width=4em, 
    text centered, rounded corners, minimum height=3em]
\tikzstyle{proof block} = [rectangle, draw, fill=magenta!10,
    minimum width=4.2em, 
     rounded corners, minimum height=1.5em]
\tikzstyle{frama block} = [rectangle, draw, fill=green!20, minimum width = 6em,
  minimum height = 4em, rounded corners]
\tikzstyle{library block} = [rectangle, draw, fill=gray!20, minimum width = 4em,
  minimum height = 2em, rounded corners]
\tikzstyle{invisible} = [opacity=0] 
\tikzstyle{every picture}+=[remember picture]

\theoremstyle{plain} \theorembodyfont{\upshape}
\newtheorem{prop}[theorem]{\indent Proposition}
\newtheorem{fact}[theorem]{\indent Fact}

\begin{document}

\mainmatter

\title{From Design to Implementation: \\an Automated, Credible Autocoding Chain for Control Systems}

\titlerunning{Automated, Credible Autocoding Chain for Control Systems}

\author{Timothy Wang\inst{1} \and Romain Jobredeaux\inst{1} \and Heber Herencia\inst{2} \and Pierre-Loïc Garoche\inst{3} \and Arnaud Dieumegard\inst{4} \and  Éric Féron\inst{1} \and Marc Pantel\inst{4}}

\authorrunning{ }

\institute{ Georgia Institute of Technology,
  Atlanta, Georgia, USA \and National Institute of Aerospace, Virginia, USA \and ONERA -- The French Aerospace Lab, Toulouse, FRANCE \and ENSEEIHT, Toulouse, France}

\maketitle

\begin{abstract}
This article describes a fully automated, credible autocoding chain for control
systems. The framework generates code, along with guarantees of high level
functional properties which can be independently verified. 
It relies on domain specific knowledge
and fomal analysis to address a context of heightened
safety requirements for critical embedded systems and ever-increasing costs of
verification and validation. The platform strives to bridge the semantic gap
between domain expert and code verification expert.
First, a graphical dataflow language is extended
with annotation symbols enabling the control engineer to express high level
properties of its control law within the framework of a familiar language.
An existing autocoder is enhanced to both generate the code implementing
the initial design, but also to carry high level properties down to
annotations at the level of the code. Finally, using customized code analysis 
tools, certificates are generated
which guarantee the correctness of the annotations with respect to the code,
and can be verified using existing static analysis tools. While only a subset of 
properties and controllers are handled at this point, the approach appears readily
extendable to a broader array of both. 
\end{abstract}

\section*{}

A wide range of today's real-time embedded systems, especially their most
critical parts, relies on a control-command computation core. The
control-command of an aircraft, a satellite, a car engine, is processed into a
global loop repeated forever, or at least during the activity of the
controlled device. This loop models the acquisition of new input values via
sensors: either from environment mesures (wind speed, acceleration, engine RPM,
\ldots) or from the human feedback via the brakes, the accelerator, the stick
or wheel control.

The cost of failure of such systems is tremendous, and examples of such failures
abound, in spite of increasingly high certification requirements. Current analysis
tools focus mainly on simulations. One obvious shortcoming is the impossiblity
to simulate all the possible scenarios the system will be subject to. More advanced
tools include static analysis modules, which derive properties of the system by formally analyzing
its semantics. However, in the specific case of control systems, analyzing the computational
core can prove arduous for these tools, whereas the engineers who designed the controller
have a variety of mathematical results which can greatly facilitate said analysis,
and evince more subtle properties of the implemented controller.

This article, following previous efforts aimed at demonstrating how control-system domain
knowledge can be leveraged for code analysis~\cite{feron:csm}~\cite{heber}, attempts
 to describe a practical implementation of a fully automated framework, which enables a control
 theorist to use familiar tools to generate credible code, that is, code delivered
with a certificate ensuring certain properties will hold on all executions. 

This article focuses on a specific class of controllers and properties in order to achieve full automation,
 but also explores various possible extensions.

The structure of the article is as follows: We first present a high level view of the general framework in Section~\ref{sec:problem}. 
We then proceed to describe how control semantics can be expressed at different levels of design, in Section 
~\ref{sec:control_semantics}. Section~\ref{sec:autocoding} describes the translation process by which graphical 
synchronous languages familiar to the control theorist can be turned into credible code. Section~\ref{sec:autoverif} 
demonstrates how a proof of correctness can be automatically extracted from the generated code.

\section{Framework and the Running example}
\label{sec:problem}
\begin{figure}[htp]
 \includegraphics[scale=0.38]{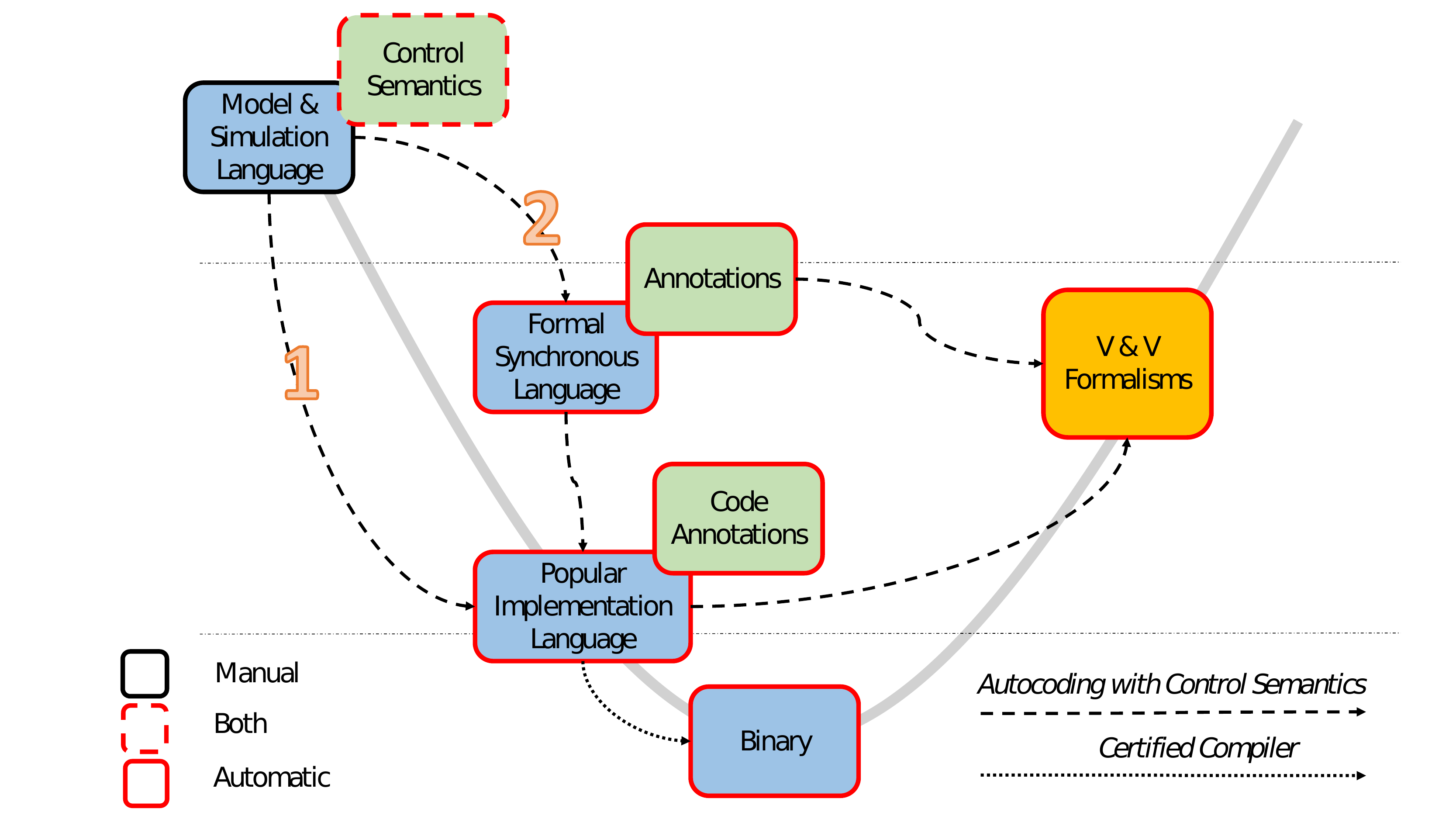}
 \caption{Automated Credible Autocoding/Compilation Chain for Control Systems}
 \label{big_picture}
\end{figure}
The framework of credible autocoding of control software using 
control semantics, which led to the development of our prototype, is summarized in 
Figure~\ref{big_picture}. 
This novel framework represents a possible conduit that will allow
the domain expert e.g. the control engineer to more efficiently produce code with
automatically certifiable safety and high-level functional properties. 
Compared to the typical model-based development paradigm, the only 
additional requirement on the control engineer that our framework stipulates is the need for
the proofs of high-level control system properties such as Lyapunov stability, 
vector margins, and other performance measures to be provided 
in the high-level specifications of the control system. 

The process of generating the proofs can be automated
using techniques from the robust control literature.
See works such as~\cite{yakubovich71s},~\cite{yakubovich62sm},~\cite{feron:csm}, or~\cite{megretskitac97}. 

\begin{figure}
	\includegraphics[scale=0.38]{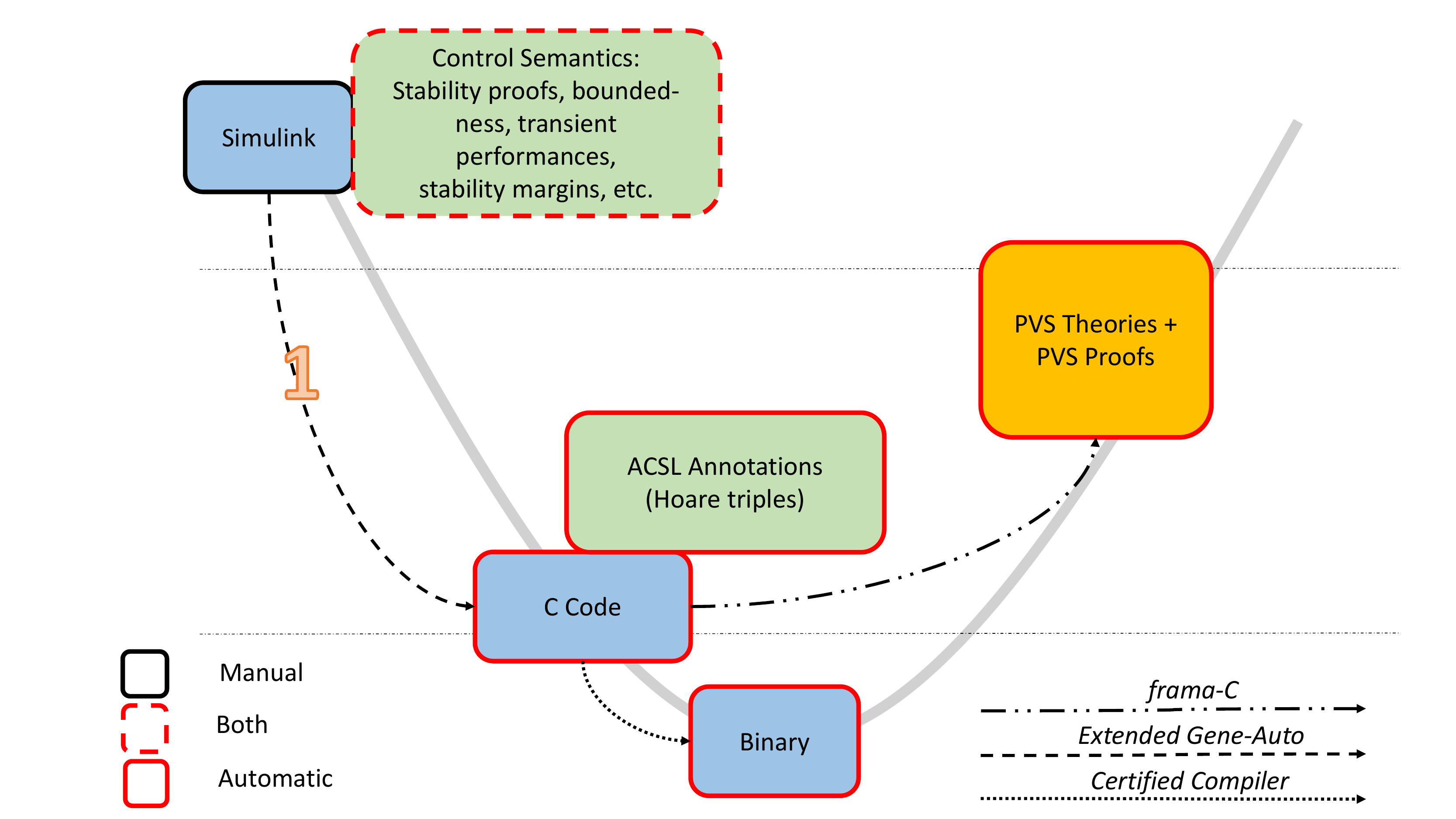}
	\caption{Autocoding with Control Semantics from Simulink to C}
	\label{ver}
\end{figure}

The are two major branches in the framework.
In the first branch, the input model along with its control semantics, is directly translated 
to the annotated source code.
On the source code level, the annotations are further translated into proof obligations, which are then discharged 
by a software theorem prover. 
To demonstrate an automation of the first branch, we built a set of specialized prototype tools, which include
an autocoder that generates the control semantics, and an annotation checker that is tailored for verifying the 
ellipsoid-based annotations on the code. 
In the second branch of the framework, 
the input model is first translated into a formal
synchronous dataflow language such as Lustre. 
At the Lustre level,  the control semantics now expressed in a new Lustre
specification language, can be analyzed and verified by adapting existing tools such as Kind 
or the one described in~\cite{prhscc12}, 

The first branch is described in more details in Figure~\ref{ver}. 
It is consists of a series of autocoding steps that translate the control semantics, inserted at the model-level
into a set of Hoare logic annotations for the output code.
The language used for the input to the framework should be a convenient 
graphical dataflow modeling language 
such as Simulink for example. 
The exact choice for the input language is up to the domain experts' preference and 
does not affect the utility of the framework as it can be adapted to other modeling languages.  
Likewise, for the output language, the choice is likely to depend on 
the preferences of the industry and the certification authority rather than 
technical reasons arising from the framework. 
For the prototype described in this paper, 
the output language was chosen to be C because of its industrial popularity 
and the wide availability of static analyzers tailored for C code.

The set of annotations in the output source code contains both the functional properties inserted by the domain expert 
and the proofs that can be used to automatically prove these properties. 
For the analysis of the annotated output, we built a prototype annotation checker that is based on the static analyzer
frama-C and the theorem prover PVS. 
For automating the proof-checking of the annotated output, 
a set of linear algebra definitions and theories were integrated into the standard NASA PVS library. 

In this paper, the 
fully automated process from the input model to the verified output is showcased for the property of
open loop stability, but the expression of other functional properties 
on the model are also discussed in this paper. 
At this point, we restrict the input to only linear controllers with possible saturations in the loop. 
The running example that we will refer to repeatedly in this paper
is the system described by the state-space difference equation in 
(~\ref{running_example}). This example has enough complexity to be representative of most
controllers used in the industry, and is simple enough such that we
can show in this paper, the output annotated code. 
The system has the states $x \in \R^{2}$, the input $y \in \R$, output $u \in \R$ and the state-transition function 
parameterized by the $4$ matrices in (~\ref{running_example}). 
\begin{example}
 \be
\ba{c} 
x_{+} = \lb \ba{cc}0.4990 & -0.05 \\0.01 & 1 \ea \rb x + \lb \ba{c} 0 \\ 0.01 \ea \rb y \\
u=\lb\ba{cc} 564.48 & 0 \ea \rb x + \lb \ba{c} 1280 \ea \rb y. 
\ea
\label{running_example} 
\ee
\label{example:main}
\end{example}
However note that
the framework has been applied to larger systems, which include the Quanser 3-degree-of-freedom Helicopter and 
an industrial F/A-18 UAV controller system. 

\section{Control Semantics}
\label{sec:control_semantics}

The set of control semantics that we can express on the model include but is not
limited to stability and boundedness.
Note that in this paper, the only control semantics of example
\ref{example:main} that we are going to demonstrate through the entire
 autocoding and proof-checking process is the proof of open-loop stability.
Nevertheless, we will also describe some other control semantics
that can be expressed on the input Simulink model and then translated into
C code annotations.

The types of systems in which we can express open-loop stability properties for
are not just limited to simple linear systems such as example
\ref{example:main}.
They also include certain nonlinear systems that can be modeled as linear
systems with bounded nonlinearities in the feedback loops.

\subsection{Control System Stability and Boundedness}

A linear control system such as the running example is formally a 
sextuple consisted of an input alphabet $y \in \R^{m}$, an output alphabet 
$u \in \R^{k}$,  a set of states $x \in \R^{n}$, an initial state $x(0)=x_{0}$, 
a linear state-transition function $\delta: (x,y) \ra x_{+}$ defined by the pair 
of matrices $\l A\in \R^{n \times n}, B\in \R^{n \times m}\r$, and a linear output function
$\omega : (x,y) \ra u$ defined by another pair of matrices $\l C \in \R^{k \times n}, D \in \R^{k \times m} \r$.  
In control, we simply express these sextuples using the following state-space formalism
\be
\ba{lc}
\dps x_{+} = A x + By, & x(0)=x_{0}\\
\dps u = C x + Dy. 
\ea
\label{ss01}
\ee
There are many inherent nonlinearities in a realistic control system 
such as time-delays, noise or unmodelled plant dynamics. 
Additionally, controller components such as safety limiters and anti-windup mechanisms also produce
nonlinearities.
One tractable way to handle all of these nonlinearities is to abstract them
as bounded nonlinear operators. 
For a more accurate representation of the control systems that are in operation, 
we also consider a class of nonlinear system that is consisted of a
linear system modelled by (\ref{ss01}) in feedback interconnections with a set of bounded
nonlinear operators. Let the input alphabet $w \in \R^{l}$ be such that $w=\sigma \l u, k \r$. 
Let $B_{w} \in \R^{n \times l}$, 
we have the following nonlinear state-space system 
\be
\ba{lc}
\dps x_{+} = A x + B y + B_{w} w , & x(0)=x_{0} \cr
\dps u = C x + D y. 
\ea
\label{ss02}
\ee
For systems described by (\ref{ss02}), which is inclusive of systems described by (\ref{ss01}), 
it is possible to compute in polynomial time the answer to the quadratic stabilizability problem. 
\begin{problem}
\begin{enumerate}
 \item Assume that the input $y$ is bounded i.e. without loss of generality let $\|y \|\leq 1$, and 
$w$ is a bounded nonlinear operator, 
does there exist a matrix $P \in \S^{n \times n}$, $P\succ 0$, such that the
quadratic function $q:  \ra x^{\m{T}} P x$ is non-increasing along the system
trajectories as $k \ra +\infty$?
\end{enumerate}
\label{problem:main}
\end{problem}
Problem \ref{problem:main} can be reformulated into a linear matrix inequality (LMI) problem. 
The details of such reformulations are skipped here in this paper as one can refer to, in the system and control literature, 
a large collection of works on the subject including \cite{yakubovich62sm}, \cite{yakubovich71s}, and \cite{boyd:lmi}. 

Without proof, here we will state that for bounded $y$ and $w=0$, the following result on an invariant for 
the system in (\ref{ss01}) holds.  
\begin{prop}
Assume $y^{\m{T}} y \leq 1$. If there exist a $P\succ 0$, $\alpha>0$, such that 
\be
\dps \lb \ba{cc} A'^{\m{T}} P A - (1-\alpha P) & A^{\m{T}} P B  \cr B^{\m{T}} P A & B^{\m{T}} P B - \alpha I_{m \times m}  \ea \rb \prec 0 
\label{lmi:main}
\ee
then $\lc x | x^{\m{T}} P x \leq 1 \rc$ is an invariant for (\ref{ss01}). 
\end{prop}
The linear matrix inequailty in (\ref{lmi:main}) can be solved for $P\succ 0$ using existing semi-definite programming solvers 
such as SeDuMi, SDPT3, CSDP, etc. 

For $w\neq 0$, we need to first characterize the nonlinearity $w=\sigma(u,t)$.  
For example, a saturation operator on the output $u$ can be captured in a sector inequality defined by $m_1, m_2>0$ and 
$\l w - m_1 u \r^{\m{T}} \l w - m_2 u \r$. 

\begin{prop}

\end{prop}

It is not possible to construct a single algorithm to automate this step in the analysis.

\subsection{Control Semantics in Simulink}

Both boundedness and stability can be expressed using a synchronous observer
with inputs $x_{i},i=1,\ldots,n$, and the boolean-valued function
\be
\dps x \ra \sum_{i,j=1,\ldots,n} x_{i} P_{ij} x_{j} \leq 1.
\label{ellipsoid_semantics}
\ee
This synchronous observer is parameterized by a symmetric matrix $P$ and a multiplier $\mu$. 
We make an important distinction between two types of ellipsoid observers. 
One is inductive and the other one is assertive. 
This distinction originates from the method used to obtain the parameter $P$ and 
is determined by the memory characteristics of the block's inputs. 
The former type must have input signals with memories i.e. if the input signals are
connected with unit delay blocks.
The inductive type expresses an invariant property of the control system loop. 
Its parameter $P$ is obtained by solving the quadratic stability problem in \ref{problem:main}.
The assertive type accepts only input signals with no memory i.e. not connected with any unit
delay block. 
Its parameter $P$ is obtained from assertions i.e. assumptions about the sizes of its input signals.

The inductive ellipsoid observer provides both the boundedness of the controller
 states and the proof of the boundedness itself.
The bound on the states can be extracted from the parameter $P$ by computing the
interval
$\dps \lb -\frac{1}{\sqrt{\sigma_{i} \l P \r }},
\frac{1}{\sqrt{\sigma_{i} \l P \r }}\rb$, where
$\sigma_{i} \l P \r$ is the $i$-th singular value of $P$.
The proof comes from the fact that the parameter $P$ is the
answer to the quadratic stability problem hence
the following holds true.
\begin{fact}
$\forall x \in \R^{n}$, if $y$ is bounded and $P$ is the solution to the
quadratic stability problem in \ref{problem:main}, then
$\l A x + By \r^{\m{T}} P
\l Ax + By \r \leq x^{\m{T}} P x $.
\label{fact_stable}
\end{fact}
Fact \ref{fact_stable} is true by the construction of $P$.

\begin{figure}
\centering
\includegraphics[scale=0.6]{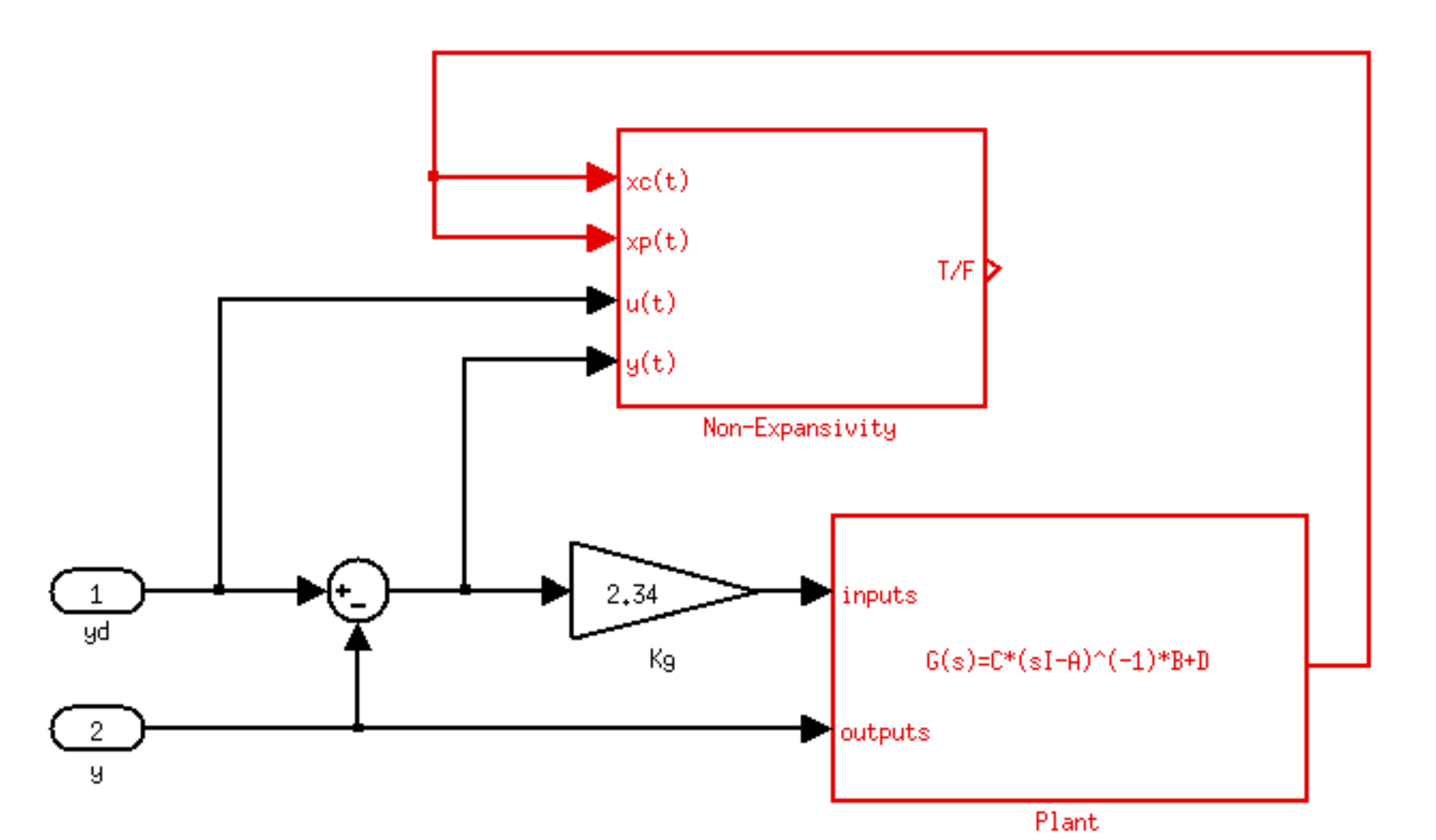}
\caption{Expressing Vector Margin of the Closed-Loop System}
\label{closed_loop}
\end{figure}

For expressing the ellipsoid observers on the Simulink model, we constructed a
custom S block denoted as \emph{Ellipsoid} to represent the ellipsoid observer.
Additionally, for expressing the operational semantics of the plant,
we constructed a custom S block denoted as \emph{Plant}.
Its semantics is similar to Simulink's discrete-time state-space block with two key differences.
One is that the input to the \emph{Plant} block contains both the input and output of the plant.
The other is that the output from the \emph{Plant} block are the internal states of the plant. 
Note that the outputs from the \emph{Plant} block represent variables with memories so they can 
also be inputs to an inductive ellipsoid observer. 
Other properties can also be expressed such as non-expansivity from the dissipativity framework. 
The \emph{Non-Expansivity} block, when connected with 
the appropriate inputs and outputs, can be used to express a variety of performance measures such as 
the $H_{\infty}$ characteristic of the system or the closed-loop vector margin of the 
control system. An example of such usage is shown in figure \ref{closed_loop} where the 
closed-loop vector margin of a constant gain controller is expressed using a combination of the
\emph{Plant} block and the \emph{Non-Expansivity} block. 

In this paper, we focus on the current fully-automated treatment of 
the open-loop stability properties, hence we will not consider
the semantics displayed in figure (\ref{closed_loop}) beyond the
description here.

\begin{figure}[htp]
 \centering
 \includegraphics[scale=0.5]{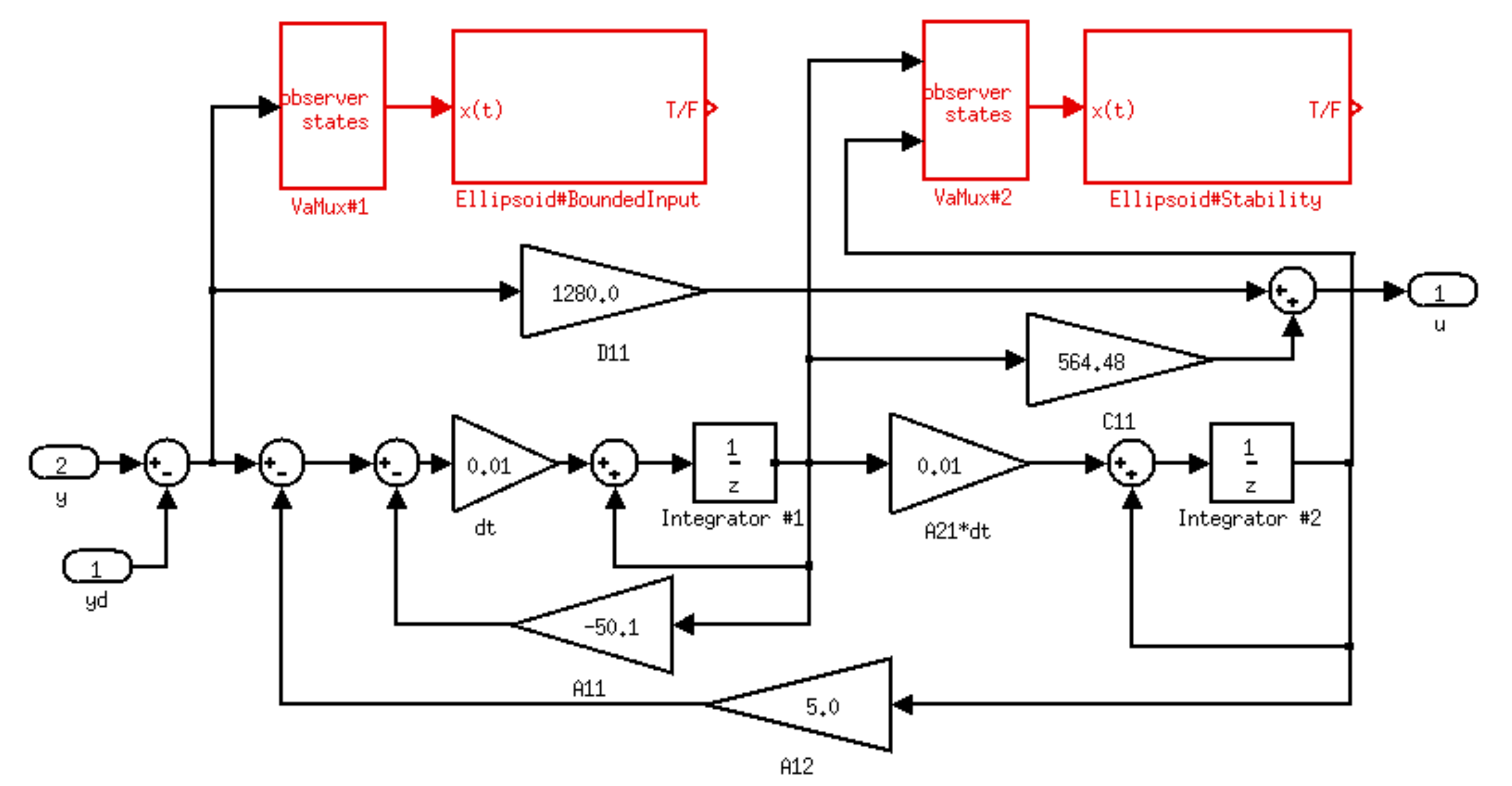}
 \caption{Running Example with Synchronous Observers}
 \label{2dss}
\end{figure}

For the running example, we have a Simulink model connected with
with two synchronous observers.   
The observers are displayed in red for clarity's purpose. 
In the Simulink model, we introduce an error into the control system by flipping the sign of gain block
\emph{A11} in figure \ref{2dss}. This error will be referred to again in the next section of paper. 
We made the following assumption for the quantity $y-y_{d}$: 
\be
\dps \| y - y_{d} \|\leq 0.5, 
\label{elp01}
\ee
which is expressed in figure \ref{2dss} by the Ellipsoid block \emph{BoundedInput} with the parameters $P=0.5$ and multiplier $\mu=0.0009$. 
The stability proof is expressed in figure \ref{2dss} by the Ellipsoid block \emph{Stability} with the parameters
$\dps P=\lb \ba{cc} 6.742\times 10^{-4} & 4.28\times 10^{-5} \cr 4.28\times 10^{-5} & 2.4651\times 10^{-3} \ea \rb$ and $\mu=0.9991$. 
The observer blocks in figure \ref{2dss} are connected to the model using the \emph{VaMux} block. 
The role of the \emph{VaMux} block is to concatenate a set of scalar signal inputs into a single vector output.  
This special block was constructed because the \emph{Ellipsoid} observer block can accept only a single vector input. 

\subsection{Control semantics at the level of the C code}
For the specific problem of open loop stability, the expressiveness needed at
the C code level is twofold. On the one hand, a means of expressing that a
vector composed of program variables is a member of an ellipsoid is required.
This entails a number of underlying linear algebra concepts. On the other hand,
a eans of providing the static analysis tools with indications on how to
proceed with the proof of correctness.

ACSL, the ANSI/ISO C Specification Language, is an annotation language for C
\cite{baudin:acsl}.It is expressive enough, and
its associated verification tool, Frama-C, offers a wide variety of
backend provers which can be used to establish the correctness of the annotated
 code.
\subsubsection{Linear Algebra in ACSL}
A library of ACSL symbols has been developed to express concepts and properties
pertaining to linear algebra. In particular, types have been defined for
matrices and vectors, and predicates expressing that a vector of variables is a
member of the ellipsoid $\mathcal{E}_P$ defined by
$\{x\in \mathbb{R}^n: x^{\mathrm{T}}Px\leq 1\}$, or the ellipsoid
$\mathcal{G}_X$ defined by
$\left \{x\in \mathbb{R}^n: \left [
\begin{array}{cc}
1 & x^{\mathrm{T}} \\
x & X
\end{array} \right ] \geq 0 \right\}$. For example, expressing that the vector
composed of program variables $v_1$ and $v_2$ is in the set $\mathcal{E}_P$
where $P=\begin{pmatrix}1.53 & 10.0 \\ 10.0 & 507 \end{pmatrix}$, can be done
with the following ACSL code:
\begin{tikzboxtt}{ACSL}
/*@ logic matrix P = mat_of_2x2_scalar(1.53,10.0,10.0,507);
  @ assert in_ellipsoid(P,vect_of_2_scalar(v_1,v_2)); */
\end{tikzboxtt}
The stability of ellipsoid $\mathcal{E}_P$ throughout any program execution
can be expressed by the following \emph{loop invariant}:
\begin{tikzboxtt}{ACSL+C}
//@ loop invariant in_ellipsoid(P,vect_of_2_scalar(v_1,v_2));
while (true)\{
  //loop body
\}
\end{tikzboxtt}
In terms of expressiveness, this latter annotation is all that is required to
express open loop stability of a linear controller. However, in order to
facilitate the proof, intermediate annotations are added within the loop to
propagate the ellipsoid through the different variable assignments, as suggested
in \cite{feron:csm} and expanded on in section \ref{sec:autocoding}. For this
reason, a loop body instruction can be annotated with a local contract, like so:
\begin{tikzboxtt}{ACSL+C}
/*@ requires in_ellipsoid(P,vect_of_2_scalar(v_1,v_2));
  @ ensures in_ellipsoid(Q,vect_of_3_scalar(v_1,v_2,v_3));*/
\{
// assignment of v_3
\}
\end{tikzboxtt}
\subsubsection{Including proof elements}
An extension to ACSL, as well as a plugin to Frama-C have been developed.
They make it possible to indicate the proof steps needed to show the correctness
of a contract, by adding extra annotations. For example, the following syntax:

\begin{tikzboxtt}{ACSL+C}
/*@ requires in_ellipsoid(P,vect_of_2_scalar(v_1,v_2));
  @ ensures in_ellipsoid(Q,vect_of_3_scalar(v_1,v_2,v_3));*/
  @ PROOF_TACTIC (use_strategy (Intuition));
\{
// assignment of v_3
\}
\end{tikzboxtt}
signals Frama-C to use the strategy 'Intuition' to prove the correctness of the
local contract considered. Section \ref{sec:autoverif} expands on this topic.

\subsection{Control semantics in PVS}
Through a process described in section \ref{sec:autoverif}, verifying the
correctness of the annotated C code is done with the help of the interactive
 theorem prover PVS. This type of prover normally rely on a human in the loop
 to provide the basic steps required to prove a theorem. In order to reason
about control systems, linear algebra theories have been developed. General
properties of vectors and matrices, as well as theorems specific to this
endeavor have been written and proven manually within the PVS environment.
\subsubsection{Basic types and theories}
Introduced in \cite{heber} and available at \todo{URL} as part of the larger
NASA PVS library, the PVS linear algebra library enables one to reason about
matrix and vector quantities, by defining relevant types, operators and
predicates, and proving major properties. To name a few:
\begin{itemize}
\item A vector type.
\item A matrix type, along with all operations relative to the algebra of
matrices.
\item Various matrix subtypes such as square, symmetric and positive definite
matrices.
\item Block matrices
\item Determinants
\item High level results such as the link between Schur's complement and
positive definiteness
\end{itemize}
\subsubsection{Theorems specific to control theory}
In \cite{heber}, a theorem was introduced, named the ellipsoid theorem.
A stronger version of this theorem, along with a couple other useful results in
proving open loop stability of a controller, have been added to the library.
The following theorem

\begin{tikzboxtt}{PVS}
ellipsoid_general: LEMMA
 \(\forall\) (n:posnat,m:posnat, Q:SquareMat(n),
           M: Mat(m,n), x:Vector[n], y:Vector[m]):
              in_ellipsoid_Q?(n,Q,x)
              AND y = M*x
        IMPLIES
        in_ellipsoid_Q?(m,M*Q*transpose(M),y)
\end{tikzboxtt}
expresses in the PVS syntax how a generic ellipsoid $\mathcal{G}_Q$ is
transformed into $\mathcal{G}_{MQM^T}$ by the linear mapping $ x\mapsto Mx$.
This next theorem:

\begin{tikzboxtt}{PVS}
ellipsoid_combination: LEMMA
 \(\forall\) (n,m:posnat, lambda_1, lambda_2: posreal, Q_1: Mat(n,n),
           Q_2: Mat(m,m), x:Vector[n], y:Vector[m], z:Vector[m+n]):
              in_ellipsoid_Q?(n,Q_1,x)
              AND in_ellipsoid_Q?(m,Q_2,y)
              AND lambda_1+ lambda_2 = 1
              AND z = Block2V(V2Block(n,m)(x,y))
        IMPLIES
        in_ellipsoid_Q?(n+m,Block2M(M2Block(n,m,n,m)(1/lambda_1*Q_1,
                       Zero_mat(m,n),Zero_mat(n,m),1/lambda_2*Q_2)),z)
\end{tikzboxtt}
expresses how, given 2 vectors $x$ and $y$ in 2 ellipsoids
$\mathcal{G}_{Q_1}$ and $\mathcal{G}_{Q_2}$, and a multiplier $\mu$,
it can always be said that
$\begin{pmatrix} x \\ y \end {pmatrix} \in \mathcal{G}_Q$, where
$Q=
\begin{pmatrix}
\frac{Q_1}{\mu} & 0 \\
0 & \frac{Q_2}{1-\mu}
\end{pmatrix}
$

These 2 theorems are used heavily in section \ref{sec:autoverif} to prove the correctness of a block.




\section{Autocoding with Control Semantics}
\label{sec:autocoding}

We have so far defined the annotations blocks to express control semantics at the model level. 
In this Section, we describe in more details, the prototype autocoder that we built to 
transform the set of control semantics, defined in Section (~\ref{sec:control_semantics}), 
into analyzable ACSL annotations on the code. 
The prototype is based on Gene-Auto, which is an existing industrial-capable automatic code generator 
for real-time embedded systems\cite{nassimaformal09}.

\subsection{Introduction to Gene-Auto}

\begin{figure}[htp]
\includegraphics[scale=0.4]{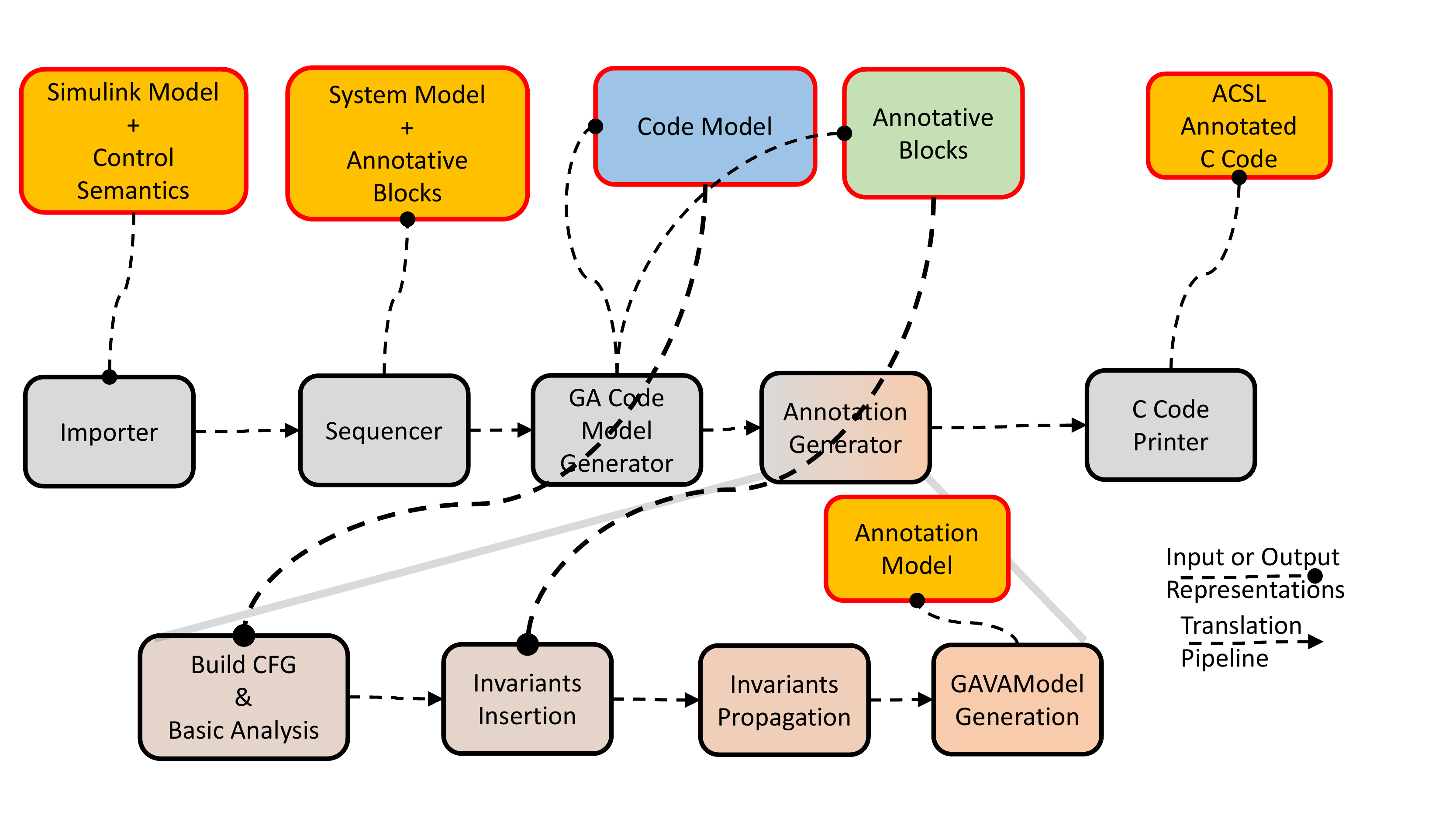} 
\caption{Gene-Auto Translation Pipeline/Modifications}
\label{geneauto}
\end{figure}

Gene-Auto's translation architecture is consisted of a sequences of independent model transformation stages. 
This classical, modular approach to code generator design has the advantage of allowing relatively easy insertion of additional
transformation and formal analysis stages such as the annotations generator in our prototype. 
Figure~\ref{geneauto} has an overview of Gene-Auto's transformation pipeline. 
The process goes through two layers of intermediate languages. The first one, called 
the \emph{GASystemModel}, is a dataflow language semantically similar to 
the discrete subset of the Simulink formalism. 
The input Simulink model, after being imported, is first transformed into the system model. 
The system model, which is expressed in the \emph{GASystemModel} language, 
is then transformed into the code model. 
The code model is in the \emph{GACodeModel} language representation, 
which is semantically similar to imperative programming languages such as C or Ada.
The key translational stages from the input model to the output code are the importer, the pre-proccessor, the block sequencer, the typer, 
the \emph{GACodeModel} generator, and finally the printer. 
For our prototype, we have recycled much of the translational stages up to 
the \emph{GACodeModel} generator. 
From this point on, the prototype tool branch off from the original pipeline until the printer stage, in which 
much of that was also recycled. 

\subsection{Annotations Generation} 

The control semantics are also first transformed
into a \emph{GASystemModel} representation. 
This transformation step is unaltered from the original Gene-Auto as the formalisms to express 
the control semantics and the controller model on the Simulink level are very similar. 
In the \emph{GACodeModel} generation stage, the system blocks that express the control semantics are 
not transformed into a \emph{GACodeModel} representation. Instead, they are imported into 
the annotations generation module. 
The annotations generation module is initiated after the code model has been constructed from the system model. 
The module is responsible for the insertion of the control semantics blocks onto the code model and then 
transforming them into a \emph{GACodeModel} Specification Language representation called the $\emph{GAVAModel}$. 
The code model with the control semantics expressed in \emph{GAVAModel} becomes the output of the 
annotations generator. 
This new representation of code and properties is dubbed as the annotation model. 

\begin{figure}
	\includegraphics[scale=0.4]{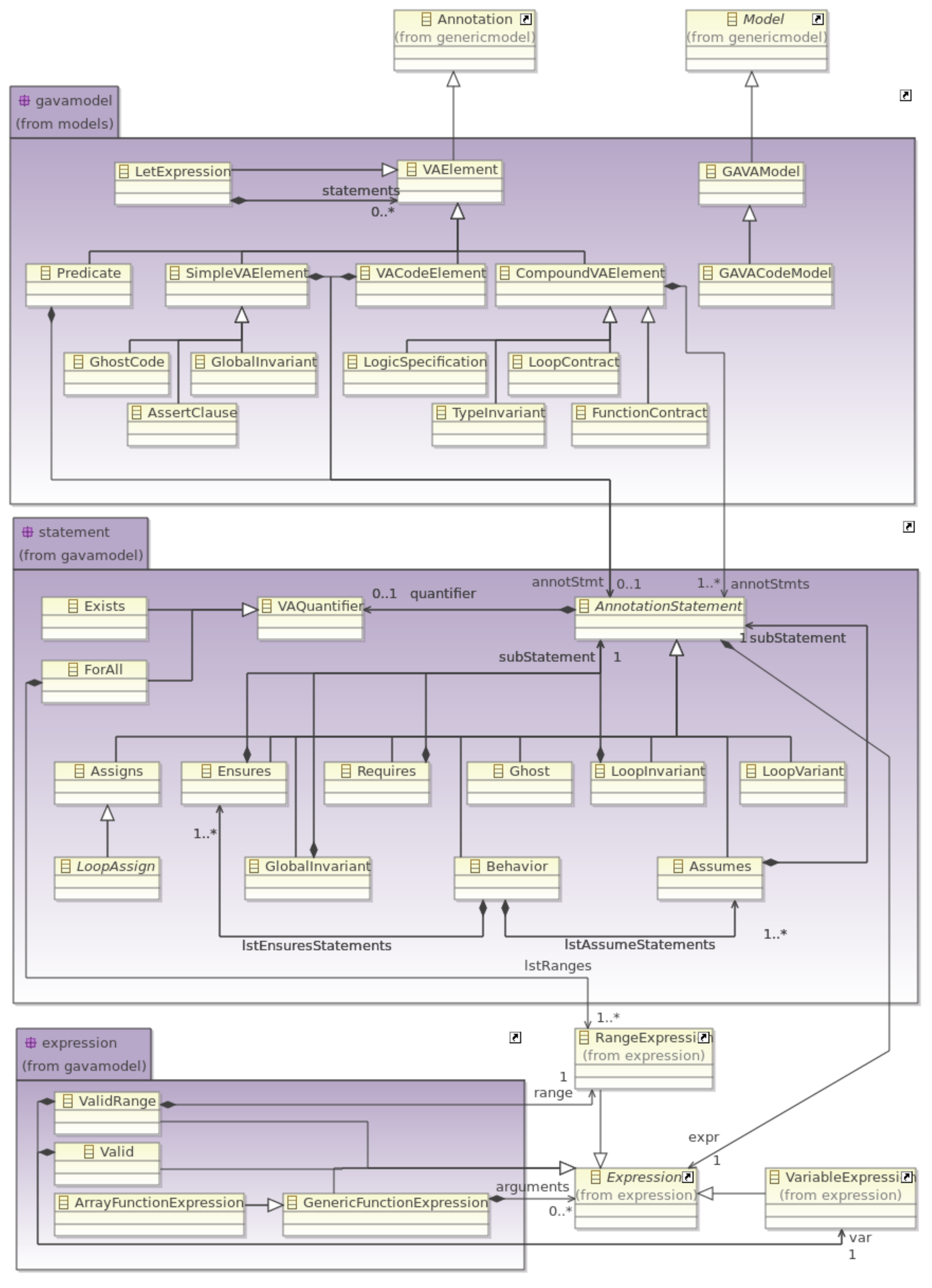}
 \caption{Specification Language for the Gene-Auto CodeModel} 
 \label{gavamodel}
\end{figure}

The \emph{GAVAModel} was added specifically for expressing the control properties and proofs.
However it is based the specification language on ACSL\cite{baudin:acsl} 
so it can be used to express other first-order properties about the generated code.  
A summary of its elements can be seen in Figure~\ref{gavamodel}. 

Following Gene-Auto's modular transformation architecture, the annotations generator is added as 
an independent module within the \emph{GACodeModel} generation module. 
The pipeline for the annotation generation module is also shown in Figure~\ref{geneauto}. 
The major stages in converting the annotative system blocks in Figure~\ref{2dss} 
to the annotation model representation are the following: 
\begin{enumerate}
 \item Importing. Convert the code model for into a control flow graph for analysis.  
 \item Analysis. Basic analysis of the code model like unrolling of finite loops, constant propagation, and affine analysis. 
	 All informations obtained about the code model during this step is stored in the control flow graph.  
 \item Plant block analysis and insertion. Process the \emph{Plant} block, and inserted it as an affine transformation object onto the control flow graph. 
    If there is no plant, this step is skipped altogether.  
 \item Observers analysis and insertion.  The \emph{Ellipsoid} observers are typed based on their inputs and then inserted onto the control flow graph 
	 as either assertive invariants or inductive ones. 
 \item Invariant propagation. Additional ellipsoids are generated from the inserted observers using 
	 basic Hoare logic with the two additions: the \emph{AffineEllipsoid} and the \emph{SProcedure} rules. 
 \item GAVAModel generation. Convert all the invariants to \emph{GAVAModel} and insert them as annotations onto the code model. 
	 This creates the annotation model. 
\end{enumerate}

Here we will skip the details about stages 1 to 2 as they were constructed using standard existing algorithms. 
We will also skip any details of stage 3 as in this paper, as there is no \emph{Plant} block in the running
example. 
While information about the plant is critical to expressing closed-loop properties such as performance margins,
the scope of this paper is focused mostly on the open-loop stability properties. 

\subsection{Observers analysis and Insertion} 
Here we will give a short description of the ellipsoid typing and insertion process.  
The open-loop control semantics are structured in such way that there is one
inductive ellipsoid observer on the model with several other assertive ellipsoid observers on 
the inputs. 

\subsubsection{Typing of the Ellipsoids}

The typing algorithm traces the input ports of the block to their corresponding
variables and then checks if all of the variables are linked directly to any port from an
Unit Delay blocks or an output port on a \emph{Plant} block.
Since this algorithm operates on the \emph{GASystemModel} level, 
the typing is done right before the transformation of 
the \emph{Ellipsoid} blocks to ellipsoid invariants.

After the conversion, the equivalent $Q$-form\footnote[1]{If matrix $P^{-1}$ exists and let $Q=P^{-1}$, then
$x^{\m{T}} P x \leq 1$ is equivalent to $\dps \lb \ba{cc} 1 & x^{\m{T}} \\ x & Q \ea \rb \geq 0$ which is the $Q$-form} is computed
	for each ellipsoid invariant.
This conversion is necessary as all subsequent transformation of the ellipsoid invariants are done 
in the $Q$-form due to the possibility of a degenerate ellipsoid. 
A degenerate ellipsoid can be visualized in two dimensions as a line segment.

\subsubsection{Insertion}
For the assertive ellipsoids, the insertion location is computed using algorithm~\ref{alg01}. 
The function LHS returns the left hand side of an assignment statement. 
The variable G is the control flow graph of the code model. 
Each node of the control flow graph contains a statement. 
Simply put, the algorithm founds the first assignment statement where the ellipsoid's state variable is 
be assigned to and inserts the ellipsoid invariant into the statement's list of post-conditions. 
\begin{algorithm}
\caption{Assertive Ellipsoid Insertion} 
\begin{algorithmic}[1]
	\Procedure {EllipsoidInsertion} {$G$,$\mathcal{E}$}
		\For {$i \leftarrow 1, G.nNodes$} 
		\If {$currentNode == AssignStatement$}
		\If {$LHS(currentNode) \in E.States$}
			\State $add(E,currentNode.PostConditions)$
			\EndIf
		\EndIf
		\EndFor
	\EndProcedure 
\end{algorithmic}
\label{alg01}
\end{algorithm}
For the running example, there is a single assertive ellipsoid invariant on the 
state $Sum4=y-y_{d}$ and that is inserted into the code as a post-condition of 
the line of code that assigns the quantity $y-yd$ to the variable \bq{Sum4}. 
This is shown in the autocoded output in Figure~\ref{ast}. 
The assertive ellipsoid is defined by the matrix variable \bq{QMat\_11} and is expressed using
the function \bq{in\_ellipsoidQ}\footnote[2]{The function \bq{in\_ellipsoidQ} defines the ellipsoid in the $Q$-form}. 
The \emph{assumes} keyword in ACSL means the formula is an assertion rather than a property to be checked.  
\begin{figure}[htp]
\centering
\begin{tikzboxtt}{ACSL+C}
\{
  Sum4 = discrete_timeg_no_plant_08b_y - discrete_timeg_no_plant_08b_yd;
\}
/*@ 
        logic matrix QMat_11 = mat_of_1x1_scalar(0.5);
*/
   /*@ 
            behavior ellipsoid8_0:
            assumes in_ellipsoidQ(QMat_11,vect_of_1_scalar(Sum4));
            ensures in_ellipsoidQ(QMat_12,vect_of_2_scalar(Sum4,D11));
            @ PROOF_TACTIC (use_strategy (AffineEllipsoid));
    */
\end{tikzboxtt}
\caption{Assertive Ellipsoid Invariant $\ra$ Assumption in ACSL} 
\label{ast}
\end{figure} 

The insertion of an inductive ellipsoid invariant is handled differently. 
The inductive ellipsoid is inserted as pre and post-conditions respectively at
the beginning and end of a function body.
It is also inserted as a pre and post-conditions on the function itself. 
The function body is obtained by searching for the function that is linked with all the states of the ellipsoid invariant. 
If the ellipsoid invariant is linked to more than one function then the search goes up the function call hierarchy until it founds a function that is 
linked with all the states of the ellipsoid invariant. 
Figure~\ref{prepost} shows a portion of the autocoded output of the running example. 
The first three ACSL annotations in Figure~\ref{prepost} defines 
the matrix variables \bq{QMat\_0}, \bq{QMat\_1} and \bq{QMat\_2}. 
All three matrix variables parameterize the same ellipsoid as the one
obtained from the stability analysis and inserted into the Simulink model as the \emph{Ellipsoid\#Stability} observer. 
Using the ACSL keywords \emph{requires} and \emph{ensures}, we can express pre and post-conditions for lines of code as well as 
functions.
The fourth ACSL annotation in Figure{prepost} expresses the inserted ellipsoid pre and post-conditions for the state-transition 
function of the controller: \bq{discrete\_timeg\_no\_plant\_08b\_compute}. 
The last ACSL annotation in Figure{prepost} contains a copy of the pre-condition from the fourth annotation. 
This is the inserted ellipsoid pre-condition for the beginning of the function body. 
\begin{figure}[htp]
\centering
\begin{tikzboxtt}{ACSL+C}
/*@ 
        logic matrix QMat_0 = 
	mat_of_2x2_scalar(1484.8760396857954,-25.780980284188082,
			 -25.780980284188082,406.11067541120576);
*/
/*@ 
        logic matrix QMat_1 = 
	mat_of_2x2_scalar(1484.8760396857954,-25.780980284188082,
			 -25.780980284188082,406.11067541120576);
*/
/*@ 
        logic matrix QMat_2 = 
	mat_of_2x2_scalar(1484.8760396857954,-25.780980284188082,
			-25.780980284188082,406.11067541120576);
*/
/*@ 
        requires in_ellipsoidQ(QMat_0,
	vect_of_2_scalar(_state_->Integrator_1_memory,
			_state_->Integrator_2_memory));
        requires \\valid(_io_) && \\valid(_state_);
        ensures in_ellipsoidQ(QMat_1,
	vect_of_2_scalar(_state_->Integrator_1_memory,
	_state_->Integrator_2_memory));
*/
void discrete_timeg_no_plant_08b_compute(
		t_discrete_timeg_no_plant_08b_io *_io_, 
		t_discrete_timeg_no_plant_08b_state *_state_)\{
	.
	.
	.
	.
	.

      /*@ 
            behavior ellipsoid0_0:
            requires in_ellipsoidQ(QMat_2,
	    vect_of_2_scalar(_state_->Integrator_1_memory,
	    _state_->Integrator_2_memory));
            ensures in_ellipsoidQ(QMat_3,
	    vect_of_3_scalar(_state_->Integrator_1_memory,
	    _state_->Integrator_2_memory,x1));
            @ PROOF_TACTIC (use_strategy (AffineEllipsoid));
    */
\{
    x1 = _state_->Integrator_1_memory;
\}
\end{tikzboxtt}
\caption{Inductive Ellipsoid Invariant $\ra$ Pre and Post-conditions of the State-Transition Function} 
\label{prepost}
\end{figure}

\subsection{Invariant Propagation} 

The manual forward propagation of ellipsoid invariants was described in \cite{feron:csm}. 
Here we will describe the automatic version of the same process as it is implemented in the prototype. 
First we give a brief introduction of Hoare logic followed by a description of a
targeted extension of Hoare logic to ellipsoid invariants in linear systems.   
Finally we demonstrate this process on the running example.  

\subsubsection{Hoare Logic}

Hoare Logic is a formal proof system that is used to reason about assertions and properties of a program. 
The key element in Hoare logic is the Hoare triple, which is consisted of a boolean-valued formula, 
followed by a line of code and then followed by another boolean-valued formula. 
\be
\dps \lc a_{1} \rc \mathcal{C}_{1} \lc a_{2} \r 
\label{hoare_triple}
\ee
The meaning of the Hoare triple in (~\ref{hoare_triple}) is as follows: if formula $a_{1}$ is true, 
then after the execution of the line of code $\mathcal{C}_{1}$, the formula $a_{2}$ is also true.
In addition to the Hoare triple, a set of axioms and inferences rules were constructed to
reason about the axiomatic semantics of the entire program.  
Two examples of such inference rules for a simple imperative language from \cite{hoareaxiom69} 
are listed in Table~\ref{hoare_axioms}. 
In order to annotate every lines of the generated code with ellipsoidal pre and post-conditions, 
we also constructed two specialized inference rules for ellipsoids. 
The first rule is derived from the property of ellipsoid transformation under linear mapping. 
The second rule is derived from the well-known S-Procedure method in control theory. 
These two inference rules are sound in the real number domain. 
\begin{table}
\centering
 \begin{tabular}{c}
    $\frac{\lc \mathcal{E}_{1}\rc P_{1} \lc \mathcal{E}_{2} \rc
    \lc \mathcal{E}_{2} \rc P_{2} \lc \mathcal{E}_{3} \rc}{	
    \lc \mathcal{E}_{1} \rc P_{1};\rm{  }P_{2} \lc \mathcal{E}_{3} \rc }$(\bq{Composition})  \\ 
    \\
    $\frac{\vDash \l \mathcal{E}_{0} \Ra \mathcal{E}_{1}\r \lc \mathcal{E}_{1} \rc P \lc \mathcal{E}_{2} \rc
    \vDash \l  \mathcal{E}_{2} \Ra \mathcal{E}_{n}\r}{\lc \mathcal{E}_{0} \rc P \lc \mathcal{E}_{n} \rc}$ 
    (\bq{Consequence})\\
    \\
 \end{tabular}
\caption{Rules in Hoare Logic} 
\label{hoare_axioms}
\end{table}

\subsubsection{Affine Transformation} 
For the linear propagation of ellipsoids, we define the \emph{AffineEllipsoid} rule. 
This rule is applied when the line of 
code is an assignment statement and has a left-hand expression that is
linear. 
Let the expression $a$ be such that $\dps \Lb a \Rb \vDash L y $, where $y \in \R^{m}$ is vector of 
program states and $L \in \R^{1 \times m}$. 
We define the schur form of an ellipsoid using the function 
$S: (Q,x)\ra \dps \lb \ba{cc} Q & x^{\m{T}} \cr x & 1 \ea \rb $ and $\mathcal{Q}_{1,x}=\lc x | S(Q_{1},x)>0\rc$. 
For the Hoare triple, 
\be
\dps \lc \mathcal{Q}_{1}(x) \rc \emph{z:=a} \lc \mathcal{Q}_{2}(x \cup z) \rc, 
\label{hoare_triple_ellipsoid}
\ee where $\mathcal{Q}_{n}(x) := \lc x | S(Q_{n},x)>0\rc$,
we want to generate $\mathcal{Q}_{2}$ to satify the partial correctness of 
(~\ref{hoare_triple_ellipsoid}), we use the following rule.
\begin{figure}
 \be\ba{c}
\dps \frac{}{\lc \mathcal{Q}_{n}(x) \rc z:=a \lc \mathcal{Q}_{n+1} (x \cup z)\rc}, 
\dps Q_{n+1}=\mathcal{F} \l Q_{n},\psi(L,y,x),\phi(z,x) \r
 \ea
 \ee
 \caption{AffineEllipsoid}
 \label{linear} 
\end{figure}

The function $\mathcal{F}$ is defined as follows: given the functions $\psi: (L,y,x)\ra \R^{1 \times n}$ 
and $\phi: (z,x) \ra \Z$, we have 
\be
\ba{c}
\dps \mathcal{F} :\l Q_{n},\psi(L,y,x),\phi(z,x) \r \ra T\l \psi(L,y,x),\phi(z,x) \r^{\m{T}} Q_{n} 
T\l \psi(L,y,x),\phi(z,x) \r \cr
\dps T\l\psi(L,y,x),\phi(z,x)_{i,j} \r:=
\lc \ba{ll} 
1\rm{,} & 0\leq i,j \leq n \wedge i=j \wedge i\neq \phi(z,x) \cr
0\rm{,} & 0 \leq i,j\leq n\wedge i\neq j\wedge i \neq \phi(z,x)\cr
\psi(y,x)_{1,j} \rm{,}  & i=\phi(z,x)\wedge 0\leq j\leq n 
\ea \right. \cr
\dps \psi(L,y,x)_{1,j}:=\lc \ba{ll} 
\dps L(1,k)\rm{,} & 0\leq j,k\leq n \wedge x_{j} \in y \wedge y_{k} = x_{j}  \cr
\dps 0 \rm{,} & 0 \leq j \leq n \wedge x_{j} \notin y
\ea \right.\cr
\dps \phi(z,x):=\lc \ba{ll} 
i\rm{,} & z \in x \wedge z=x_{i}\cr
n+1\rm{,} & z\notin x 
\ea \right.
\ea
\label{linear_trans}
\ee

The \emph{ReduceEllipsoid} rule is related to the \emph{AffineEllipsoid} hence they are grouped together. 
We have the following rule, 
\begin{figure}
 \be\ba{c}
\dps \frac{}{\lc \mathcal{Q}_{n}(x) \rc P \lc \mathcal{Q}_{n+1} (\lc x_{i}\rc \setminus \lc z \rc)\rc}, 
\dps Q_{n+1}=\mathcal{G} \l Q_{n},\theta(z,x) \r
 \ea
 \ee
 \caption{ReduceEllipsoid}
 \label{reduce} 
\end{figure}

The function $\mathcal{G}$ is defined as the following: given the function $\theta: (z,x) \ra \Z$, we have
\be
\ba{c}
\dps \mathcal{G} :\l Q_{n},\theta(z,x) \r \ra T\l \theta(z,x) \r^{\m{T}} Q_{n} 
T\l \theta(z,x) \r \cr
\dps T\l \theta(z,x)_{i,j} \r:=
\lc \ba{ll} 
1\rm{,} & 0\leq i,j\leq n-1 \wedge ((i<\theta(z,x) \wedge i=j) \vee (i \geq \theta(z,x) \wedge j=i+1)) \cr
0\rm{,} & 0 \leq i,j\leq n-1 \wedge ((i<\theta(z,x) \wedge i\neq j) \vee (i \geq \theta(z,x) \wedge j\neq i+1))  \cr
\ea \right. \cr
\dps \theta(z,x):=\lc \ba{ll} 
i \rm{,} & z=x_{i}\cr
\ea \right.
\ea
\label{reduce_trans}
\ee
\begin{figure}[htp]
\centering
\begin{tikzboxtt}{ACSL+C}
/*@ 
        logic matrix QMat_21 = mat_mult(mat_mult(
	mat_of_6x7_scalar(1.0,0.0,0.0,0.0,0.0,0.0,0.0,0.0,1.0,0.0,0.0,0.0,
			  0.0,0.0,0.0,0.0,1.0,0.0,0.0,0.0,0.0,0.0,0.0,0.0,
			  1.0,0.0,0.0,0.0,0.0,0.0,0.0,0.0,0.0,0.0,1.0,0.0,
			  0.0,0.0,0.0,0.0,0.0,0.01),QMat_20),
			  transpose(mat_of_6x7_scalar(1.0,0.0,0.0,0.0,0.0,0.0,
			  0.0,0.0,1.0,0.0,0.0,0.0,0.0,0.0,0.0,0.0,1.0,0.0,0.0,
			  0.0,0.0,0.0,0.0,0.0,1.0,0.0,0.0,0.0,0.0,0.0,0.0,0.0,
			  0.0,0.0,1.0,0.0,0.0,0.0,0.0,0.0,0.0,0.01)));
*/
/*@ 
        logic matrix QMat_22 = mat_mult(mat_mult(
	mat_of_6x6_scalar(1.0,0.0,0.0,0.0,0.0,0.0,0.0,1.0,0.0,0.0,0.0,0.0,
			  0.0,0.0,1.0,0.0,0.0,0.0,0.0,0.0,0.0,1.0,0.0,0.0,
			  0.0,0.0,0.0,0.0,0.0,1.0,0.0,0.0,1.0,0.0,0.0,1.0),QMat_21),
		transpose(
	mat_of_6x6_scalar(1.0,0.0,0.0,0.0,0.0,0.0,0.0,1.0,0.0,0.0,0.0,0.0,
			  0.0,0.0,1.0,0.0,0.0,0.0,0.0,0.0,0.0,1.0,0.0,0.0,
			  0.0,0.0,0.0,0.0,0.0,1.0, 0.0,0.0,1.0,0.0,0.0,1.0)));
*/
    /*@ 
            behavior ellipsoid17_0:
            requires in_ellipsoidQ(QMat_21,
	    vect_of_6_scalar(_state_->Integrator_1_memory,
	    _state_->Integrator_2_memory,x1,Sum3,Sum1,dt_));
            ensures in_ellipsoidQ(QMat_22,
	    vect_of_6_scalar(_state_->Integrator_1_memory,
	    _state_->Integrator_2_memory,x1,Sum3,dt_,Sum2));
            @ PROOF_TACTIC (use_strategy (AffineEllipsoid));
    */
\{
    Sum2 = dt_ + x1;
\}
\end{tikzboxtt}
\caption{Application of \emph{AffineEllipsoid}}
\label{affine01}
\end{figure} 
Each line of code is analyzed to determine if it is an affine assignment. 
If it is an affine assignment, the transformation matrix $L$ is extracted and stored in the control flow graph. 
For example, if we have $x=y+2*z$, then the analyzer returns the transformation matrix $L=[1, 2]$. 
Given the pre-condition 
ellipsoid $\mathcal{Q}_{i} (x) $, the \emph{AffineEllipsoid} rule only applies 
when the line of code is $y:=a$ with $\Lb a \Rb = L z \wedge z \subseteq x $. 
In Figure~\ref{affine01}, we have the annotated C output generated by our prototype for example~\ref{example:main}. 
In the example, The pre-condition is the ellipsoid in the
$Q$-form defined by the variable \bq{QMat\_21}, 
and the ensuing line of code assigns the expression \bq{dt\_+x1} to the variable \bq{Sum2}. 
The affine transformation matrix is $L=[1, 1]$ and by applying the AffineEllipsoid rule,
we have the ellipsoid transformation matrix $T$ defined by 
\be
\dps 
T=\lc 
\ba{lll}
T_{ij} = 1.0, & & (i\leq 4 \wedge  i = j ) \vee ( i=6 \wedge (j=6 \vee i=6)) \vee (i=5 \wedge j=6) \cr
 T_{ij}= 0.0, & & \rm{otherwise} 
\ea\right.
\label{tf01}
\ee

\subsubsection{S-Procedure}

The \emph{SProcedure} rule is shown in the second ACSL annotation in Figure~\ref{sproc01}. 
We will discuss the autocoded output more in detail later on 
but first we give the definition of the \emph{SProcedure} rule. 
\begin{figure}[htp]
\centering
\begin{tikzboxtt}{ACSL+C}
/*@ 
        logic matrix QMat_14 = block_m(
	mat_scalar_mult(1.0009008107296566,QMat_13),
	zeros(6,2),
	zeros(2,6),
	mat_scalar_mult(1111.111111111111,QMat_12));
*/
    /*@ 
            behavior ellipsoid9_0:
            requires in_ellipsoidQ(QMat_13,
	    vect_of_6_scalar(_state_->Integrator_1_memory,
	    _state_->Integrator_2_memory,x1,C11,Integrator_2,Sum3));
            requires in_ellipsoidQ(QMat_12,vect_of_2_scalar(Sum4,D11));
            ensures in_ellipsoidQ(QMat_14,
	    vect_of_8_scalar(_state_->Integrator_1_memory,
	    _state_->Integrator_2_memory,x1,C11,Integrator_2,Sum3,Sum4,D11));
            @ PROOF_TACTIC (use_strategy (SProcedure));
    */
\{

\}
\end{tikzboxtt}
\caption{Application of the \emph{SProcedure} Rule}
\label{sproc01}
\end{figure} 

The \emph{SProcedure} rule is activated when two or more ellipsoids can be combined correctly into a single ellipsoid. 
The definition for \emph{SProcedure} is displayed in Figure~\ref{sproc01}. 
First Let function $\mathcal{H} : \R^{n_{i} \times n_{i}} \ra \R^{ Nn \times Nn}$ be the following: 
given the function $\dim : R^{n \times n } \ra n $, 
and the function $\rho : n \in \Z^{+} \ra \sum_{i=1}^{n} \dim \l Q_{i} \r $, we have
\be
\dps 
\mathcal{H}(Q_{i})(n,m) = \lc  
\ba{lll} 
Q_{i}(n-\rho \l i-1 \r, m-\rho \l  i -1 \r), & & \rho \l i -1 \r \leq n,m \leq \rho \l i \r \cr
0.0, & & \rm{otherwise}  
\ea \right. 
\label{trans_sproc}
\ee
The \emph{SProcedure} rule is: 
\begin{figure}[htp]
 \be\ba{c}
 \dps \frac{}{\lc \mathcal{Q}_{1}(x_1) \wedge \mathcal{Q}_{2} (x_2) \wedge \ldots \wedge \mathcal{Q}_{N}(x_{N}) \rc \rm{\bq{SKIP}}  \lc \mathcal{Q}_{n+1} (x_{0} \cup x_{1} \cup \ldots \cup x_{n} )\rc } \cr
 \dps Q_{n+1}=\sum_{i=1}^{N} \mu_{i} \mathcal{H} \l Q_{i}\r. 
 \ea
 \ee
 \caption{SProcedure}
 \label{sproc} 
\end{figure}

Given the pre-conditions $\lc \mathcal{Q}_{i}(x_{i})\rc$ and the code $C$ such that $\Lb C \Rb \vDash \l y:=Lz \r$, 
the \emph{SProcedure} rule is activated only 
when all the following conditions are satisfied. 
\begin{enumerate}
	\item For each $\mathcal{Q}_{i} \l x_{i} \r$, the \emph{AffineEllipsoid} rule does not apply. 
	\item For the set $\lc \mathcal{Q}_{i} \l x_{i} \r \rc, i=1,\ldots,N$, 
		$z \subseteq \dps \bigcup_{i=1}^{N} x_{i} $. 
	\item For $\mathcal{Q}_{i} \l x_{i} \r, i=1,\ldots,N$, $z \nsubseteq x_{i} \wedge z \cap x_{i} \neq \lc \varnothing \rc  $. 
\end{enumerate} 

The multipliers are computed beforehand using the S-Procedure to ensure the soundness of the rule in the real number domain. 
For the running example, we have one ellipsoid defined by the matrix variable \bq{QMat\_12} as a pre-condition in Figure~\ref{sproc01}. 
This ellipsoid is derived from the inserted
assertive ellipsoid. 
The other ellipsiod, also a pre-condition, is defined by the matrix variable \bq{QMat\_13}. 
This ellipsoid is derived from the inserted inductive ellipsoid. 
These two ellipsoids are combined to form a new ellipsoid using the \emph{SProcedure} rule as shown in 
Figure~\ref{sproc01}. 
The new ellipsoid's matrix variable \bq{QMat\_14}, is formulated using the block matrices function \bq{block\_m}. 

\subsection{Verification of the Generated Post-condition}
After the invariant propagation step, we obtain a new ellipsoid post-condition defined by the matrix variable \bq{PMat\_24}. 
It is necessary to check 
if this new post-condition implies the inserted ellipsoid post-condition defined by the matrix variable \bq{QMat\_2} in Figure~\ref{prepost}. 
Currently, we can do a numerical verification by using a cholesky decomposition based algorithm with intervals to guarantee
a bound on the floating-point computation error.
For this particular example, because of the error introduced into the model as mentioned in Section~\ref{sec:control_semantics}, we can see that the new post-condition 
does not imply the inserted post-condition. 
\begin{figure}[htp]
\centering
\begin{tikzboxtt}{ACSL+C}
/*@ 
        logic matrix QMat_24 = 
	mat_of_2x2_scalar(3353.385756854045,-36.73496680142199,
			-36.73496680142199,406.10904154688274);
*/
    /*@ 
            behavior ellipsoid19_1:
            requires in_ellipsoidQ(QMat_23,
	    vect_of_4_scalar(_state_->Integrator_1_memory,
	    _state_->Integrator_2_memory,Sum3,Sum2));
            ensures in_ellipsoidQ(QMat_24,
	    vect_of_2_scalar(_state_->Integrator_1_memory,
	    _state_->Integrator_2_memory));
            @ PROOF_TACTIC (use_strategy (AffineEllipsoid));
    */
 \{
    _state_->Integrator_1_memory = Sum2;
 \}

\}
\end{tikzboxtt}
\caption{Generated Post-condition}
\label{end01}
\end{figure}

\section{Automatic Verification of Control Semantics}
\label{sec:autoverif}

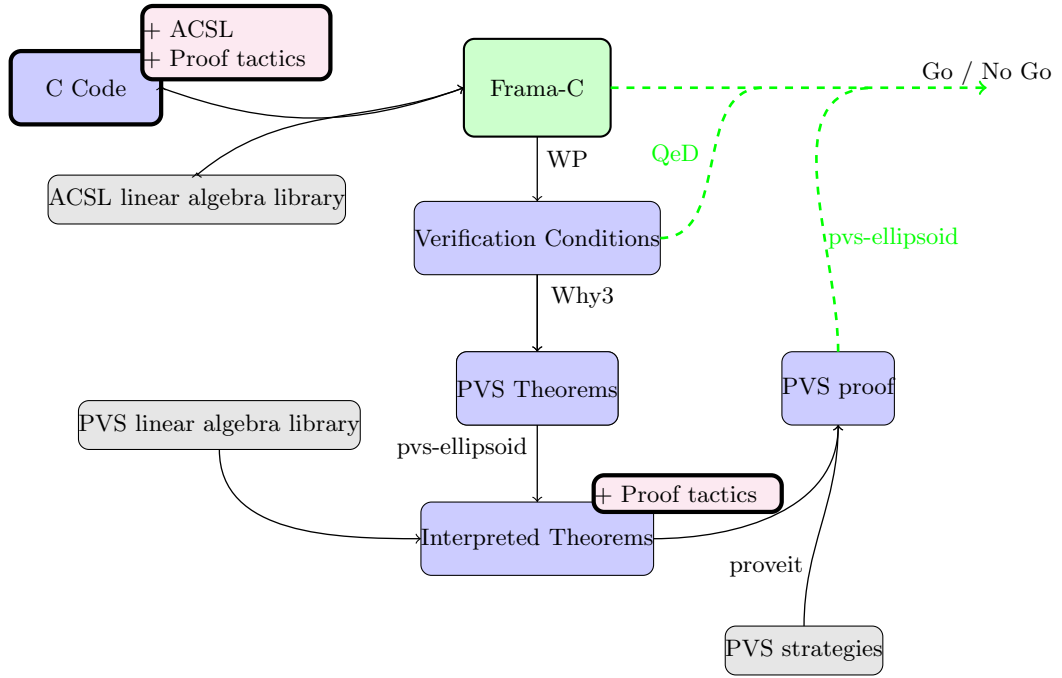
\begin{figure}
\centering
\begin{tikzpicture}[node distance = 2cm, auto, outer sep=0pt, inner sep=0pt,
   every state/.style={draw=blue!50,very
    thick,fill=blue!20}]
\node (1-8) (init_curve) {};
\node (a2) at ([shift={(-3,0)}] init_curve) {};
\node[block,ultra thick,text width=2cm, text height =] (simulink) at ([shift={(-2,1)}]a2) {C Code};
\node[proof block,ultra thick,text width=2.5cm, text height=0.8em,  minimum height=3em] (simulink_proof) at ([shift={(2,.6)}]simulink) {+ ACSL \\ + Proof tactics};
\node[frama block, thick, align=left](framablock) at ([shift={(4,1)}]a2) {Frama-C};
\draw[->,line width=.5pt] (simulink.east) edge [out=-20, in =200] (framablock.west);
\node[block] (vcs) at ([shift={(0,-2)}]framablock) {Verification Conditions};
\path[->,line width=.5pt] (framablock.south) edge (vcs.north) ;
\node (wp) at ([shift={(0.4,-.3)}]framablock.south) {WP};
\node[block] (pvsthm) at ([shift={(0,-2)}]vcs)  {PVS Theorems};
\path[->,line width=.5pt] (vcs.south) edge (pvsthm.north) ;
\node[block] (pvsthm) at ([shift={(0,-2)}]vcs)  {PVS Theorems};
\path[->,line width=.5pt] (vcs.south) edge (pvsthm.north) ;
\node (why3) at ([shift={(0.6,-.3)}]vcs.south) {Why3};
\node[block] (pvsinterp) at ([shift={(0,-2)}]pvsthm) {Interpreted Theorems};
\node[proof block,ultra thick,text width=2.5cm, text height=0.8em] (simulink_proof) at ([shift={(2,.6)}]pvsinterp) { + Proof tactics};
\path[->,line width=.5pt] (pvsthm.south) edge (pvsinterp.north) ;
\node (pvsellipsoid) at ([shift={(-1,-.3)}]pvsthm.south) {pvs-ellipsoid};
\node[library block] (pvslib) at ([shift={(-5.3,0)}]pvsthm.south east) {PVS linear algebra library};
\node[library block] (acsllib) at ([shift={(-0.3,3)}]pvslib) {ACSL linear algebra library};
\draw[->,line width=.5pt] (acsllib.north) edge [out=40, in =200] (framablock.west);
\path[->,line width=.5pt] (pvslib.south) edge [out=-90, in =180] (pvsinterp.west) ;
\node[library block] (pvsstrat) at ([shift={(2,-1)}]pvsinterp.south east) {PVS strategies};
\node[block] (pvsproof) at ([shift={(4,0)}]pvsthm) {PVS proof};
\path[->,line width=.5pt] (pvsinterp.east) edge [out=0, in =-90] (pvsproof.south) ;
\path[->,line width=.5pt] (pvsstrat.north) edge [out=90, in =-90] (pvsproof.south) ;
\node (proveit) at ([shift={(-.5,.8)}]pvsstrat.north) {proveit};
\path[->, line width=1pt, dashed, color=green] (framablock.east) edge ([shift={(5,0)}]framablock.east);
\node (intersect) at ([shift={(4.5,0)}]framablock) {};
\path[line width=1pt, dashed, color=green] (pvsproof.north) edge [out=90, in =180] (intersect);
\path[line width=1pt, dashed, color=green] (vcs.east) edge [out=0, in =180]  ([shift={(3,0)}]framablock);
\node (gonogo) at ([shift={(5,0.2)}]framablock.east) {Go / No Go};
\node[color=green] (label) at ([shift={(3.1,0)}]vcs.east) {pvs-ellipsoid};
\node[color=green] (labelqed) at ([shift={(0.2,1.1)}]vcs.east) {QeD};
\end{tikzpicture}


\caption{General view of the automated verification process. The
contribution of this Section of the article lies in the domain specific libraries
that have
been developed at the different layers of description of the code, as well as
in the generic proof strategies and the custom Frama-C plugin pvs-ellipsoid}
\label{pic_autoverif}
\end{figure}

Once the annotated C code has been generated, it remains to be proven that
the annotations are correct with respect to the code. This is achieved by checking
that each of the individual Hoare triple holds.  Figure~\ref{pic_autoverif}
presents an overview of the checking process.
First the WP plugin of Frama-C generates verification conditions for each 
Hoare triple, and discharges
the trivial ones with its internal prover QeD. Then the remaining conditions
are translated into PVS theorems for further processing, as described in
subsection~\ref{sec:c_to_pvs}. It is
then necessary to match the types and predicates introduced in ACSL to their
equivalent representation in PVS. This is done through theory interpretation~\cite{owre:sri}
and explained in subsection~\ref{sec:theory_interpretation}. Once interpreted,
the theorems can be generically proven thanks to PVS strategies, as described in
 subsection~\ref{sec:strategies}.
In order to automatize these various tasks and integrate our framework within
the Frama-C platform, which provides graphical support to display the status
of a verification condition (proved/unproved), a Frama-C plugin named
 pvs-ellipsoid, described in subsection~\ref{sec:pvs-ellipsoid}, was written.
Finally, it must be mentionned here that one last verification condition, quite
crucial, does not fall under either AffineEllipsoid of SProcedure strategies.
It is discussed in subsection~\ref{sec:lastone}

\subsection{From C code to PVS theorems}
\label{sec:c_to_pvs}
The autocoder described in the previous Section generates two C functions. One of
them is an initialization function, the other implements one execution of the
loop that acquires inputs and updates the state variables and the outputs.
It is left to the implementer to write the main function combining the two,
putting the latter into a loop, and interfacing with sensors and actuators
to provide inputs and deliver outputs. Nevertheless, the properties of open
loop stability and state-boundedness can be established by solely considering
the update function, which this Section will now focus on.
The generated function essentially follows the template shown in Figure
~\ref{pic_loop_update}.:
\begin{figure}
\centering
\begin{tikzboxtt}{ACSL+C}
/*@ requires in_ellipsoidQ(Q,vect_of_n_scalar(_state_->s_1,
                                            _state_->s_2,
                                            ...));
  @ ensures in_ellipsoidQ(Q,vect_of_n_scalar(_state_->s_1,
                                           _state_->s_2,
                                           ...));*/
void example_compute(t_example_io *_io_, t_example_state *_state_)\{
...
/*@ requires pre_i
  @ ensures post_i
  @ PROOF_TACTIC (use_strategy ( strategy_i ) )*/
\{
instruction i;
\}
...
\}
\end{tikzboxtt}
\caption{Template of the generated loop update function}
\label{pic_loop_update}
\end{figure}

Frama-C is a collaborative platform designed to analyze the source code of
software written in C. The WP plugin enables deductive verification of C
programs annotated with ACSL. For each Hoare tripe
$\{pre_i\} \mathrm{inst}_i \{post_i\}$, it generates a first order logic formula
expressing $pre_i \implies wp(\mathrm{inst}_i,post_i)$\footnote{Given a program
statement $S$ and a postcondition $Q$, $wp(S,Q)$ is
the weakest precondition on the initial state ensuring that execution of $S$
terminates in a state satisfying $Q$.}. Through the Why3
platform, these formulas can be expressed as theorems in PVS, so that, for
example, the ACSL/C triple shown in Figure~\ref{pic_acsl_triple},
taken directly from our running example, becomes the theorem shown in Figure~\ref{pic_pvs_triple}.
\begin{figure}
\centering
\begin{tikzboxtt}{ACSL+C}
/*@
requires in_ellipsoidQ(QMat_4,
                       vect_of_3_scalar(_state_->Integrator_1_memory,
                                        _state_->Integrator_2_memory,
                                        Integrator_1));
ensures in_ellipsoidQ(QMat_5,
                      vect_of_4_scalar(_state_->Integrator_1_memory,
                                       _state_->Integrator_2_memory,
                                       Integrator_1,
                                       C11));
PROOF_TACTIC (use_strategy (AffineEllipsoid));
*/
\{
    C11 = 564.48 * Integrator_1;
\}
\end{tikzboxtt}
\caption{Typical example of an ACSL Hoare Triple}
\label{pic_acsl_triple}
\end{figure}

\begin{figure}
\centering
\begin{tikzboxtt}{PVS}
wp: THEOREM
FORALL (integrator_1_0: real):
  FORALL (malloc_0: [int -> int]):
    FORALL (mflt_2: [addr -> real], mflt_1: [addr -> real],
            mflt_0: [addr -> real]):
      FORALL (io_2: addr, io_1: addr, io_0: addr, state_2: addr,
              state_1: addr, state_0: addr):
             ...
          => p_in_ellipsoidq(l_qmat_4,
                             l_vect_of_3_scalar(mflt_2(shift
                                                       (state_2, 0)),
                                                mflt_2(shift
                                                       (state_2, 1)),
                                                integrator_1_0))
          => p_in_ellipsoidq(l_qmat_5,
                             l_vect_of_4_scalar(mflt_2(shift
                                                       (state_2, 0)),
                                                mflt_2(shift
                                                       (state_2, 1)),
                                                integrator_1_0,
                                                (14112/25 *
                                                  integrator_1_0)))
\end{tikzboxtt}
\caption{Exerpt of the pvs translation of the triple shown in Figure~\ref{pic_acsl_triple}}
\label{pic_pvs_triple}
\end{figure}

Note that, for the sake of readability, part of the hypotheses of this theorem,
including hypotheses on the nature of variables, as well as hypotheses stemming
from Hoare triples present earlier in the code, are ommitted here. Note also
that in the translation process, functions like \texttt{malloc\_0} or \texttt{
mflt\_1} have appeared. They describe the memory state of the program at
different execution points.
\subsection{Theory interpretation}
\label{sec:theory_interpretation}
At the ACSL level, a minimal set of linear algebra symbols have been introduced,
along with axioms defining their semantics. Section~\ref{sec:control_semantics}
described a few of them. Each generated PVS theorem is written within a
theory that contains a translation 'as is' of these definitions and axioms,
along with some constructs specific to handling the semantics of C programs.
For example, the ACSL axiom\\
\begin{tikzboxtt}{ACSL+C}
/*@ axiom mat_of_2x2_scalar_row:
  @ {\textbackslash}forall matrix A, real x0101, x0102, x0201, x0202;
  @ A == mat_of_2x2_scalar(x0101, x0102, x0201, x0202) ==>
  @ mat_row(A) == 2; /*
\end{tikzboxtt}
becomes after translation to PVS:\\
\begin{tikzboxtt}{PVS}
 q_mat_of_2x2_scalar_row: 
  AXIOM FORALL (x0101_0:real, x0102_0:real,x0201_0:real, x0202_0:real):
           FORALL (a_0:a_matrix):
    (a_0 = l_mat_of_2x2_scalar(x0101_0, x0102_0, x0201_0, x0202_0)) =>
    (2 = l_mat_row(a_0))
\end{tikzboxtt}

In order to leverage the existing results on matrices and ellipsoids in PVS,
theory interpretation is used. It is a logical technique used to relate one
axiomatic theory to another. It is used here to map types introduced in ACSL,
such
as vectors and matrices, to the existing ones in PVS, as well as the
operations and predicates on these types. To ensure soundness, PVS requires that
what was written as axioms in the ACSL library be reproven in the interpreted
formalism.

The interpreted symbols and soundness checks are the same for each proof
objective, facilitating the mechanization of the process.
 Syntactically, a new theory is
created in which the theory interpretation is carried out, and the theorem
 to be proven is automatically rewritten by PVS in terms of its own linear
 algebra symbols. These manipulations on the generated PVS are carried out
by a frama-C plugin called pvs-ellipsoid, which will be described further in
the following subsection.
\subsection{Generically discharging the proofs in PVS}
\label{sec:strategies}

Once the theorem is in its interpreted form, all that remains to do is to apply
the proper lemma to the proper arguments. Section~\ref{sec:autocoding} described
two different types of Hoare Triple that can be generated in ACSL. Two
 pvs strategies were written to handle these possible cases. A pvs proof
strategy is a generic function describing a set of basic steps to communicate
to the interactive theorem prover in order to facilitate or even fully discharge
the proof of a lemma.

The \texttt{AffineEllipsoid} strategy handles any ellipsoid update stemming
 from a linear assignment of the variables. Recall the following theorem:
\begin{tikzboxtt}{PVS}
ellipsoid_general: LEMMA
 \(\forall\) (n:posnat,m:posnat, Q:SquareMat(n),
           M: Mat(m,n), x:Vector[n], y:Vector[m]):
              in_ellipsoid_Q?(n,Q,x)
              AND y = M*x
        IMPLIES
        in_ellipsoid_Q?(m,M*Q*transpose(M),y)
\end{tikzboxtt}
In order to apply it properly, the first step of the strategy consists of
parsing through the objectives and hypotheses of the theorem to acquire the
name and the dimensions of the relevant variables, as well as isolate the
necessary hypotheses. The second step consists of a case splitting on the
dimensions of the variable: they are given to the prover in order to complete
the main proof, and then established separately using the relevant interpreted
axioms. Next it is established that y=Mx through a tedious case decomposition
and numerous calls to relevant interpreted axioms. All the hypotheses are
then present for a direct application of the theorem. The difficulties in
proof strategy design lie in intercepting and anticipating the typecheck
constraints (tccs) that PVS introduces throughout the proof. A third
strategy was specifically written to handle them.

The S-Procedure strategy follows a very similar pattern, somewhat simpler
since the associated instruction in the Hoare triple is a skip, using 
the other main theorem presented in Section~\ref{sec:control_semantics},
\texttt{ellipsoid\_combination}.

\subsection{The pvs-ellipsoid plugin to Frama-C}
\label{sec:pvs-ellipsoid}
The pvs-ellipsoid plugin to Frama-C automatizes the steps mentionned in the
previous subsections. It calls the WP plugin on the provided C file, then, when
QeD fails to prove a step, it creates the PVS theorem for the verification
condition through Why3 and modifies the generated code to allow for theory 
interpretation. It extracts the proof tactic to be used on this specific
verification condition, and uses it to signify to the next tool in the chain,
proveit, what strategy to use to prove the theorem at hand. proveit is a command
line tool that can be called on a pvs file and attempts to prove all the
theories in there, possibly using user guidance such as the one just discussed.
When the execution of proveit terminates, a report is produced, enabling the
plugin to decide whether the verification condition is discharged or not. If
it is, a proof file is generated, enabling a replay of proof. 

\subsection{The last verification condition}
\label{sec:lastone}
As mentionned at the beginning of this Section, there is one last verification
condition that falls under neither AffineEllipsoid nor S-Procedure category.
It is the final post-condition of the main loop function contract, expressing
that the state remains in the initial ellipsoid $\mathcal{G}_P$. Through a 
number of transformations, we have a proven chain of assertions which tell
 us the state is in some ellipsoid $\mathcal{G}_P'$. The conclusion of the
 proof lies in the final test $P'-P\geq 0$. The current state of the linear
algebra library in PVS does not permit to make such a test, however a number
of very reliable external tools, like the INTLAB package of the MATLAB software
suite, can operate this check. In the case of our framework, the pvs-ellipsoid
uses custom code from~\cite{prhscc12} to ensure positive definiteness of the
matrix, with the added benefit of soundness with respect to floating point
 numbers.


\section{Related Works}
The authors would like mention and give thanks to the following related works. 
We would first like to mention Jerome Ferret and his work on the static analysis of digital filters in~\cite{ferret01}. 
It was this work that springed the connections made between the control-theoretic techniques and software analysis methods 
in~\cite{feron:csm}. 
Furthermore, we would like to mention a parallel work done by Pierre-Loic Garoche and Pierre Roux in~\cite{prhscc12} 
where policy interation is used to generate and refine ellipsoid invariants. 
We would like to give thanks to Eric Goubault and Sylvie Putot for the discussions, and mention their work on zonotopal domain
for static analyzers~\cite{goub01}.  
Finally we would like to point out Ursual Martin and her team's work on the Hoare logic for 
linear systems~\cite{ursula01}. 

\section{Conclusion}

The prototype tools and various examples described in this paper can be found
 on our \href{https://cavale.enseeiht.fr/svn/autocoding/}{svn server}. We have
demonstrated in this paper a set of prototype tools that is capable of migrating
high-level functional properties of control systems down to the code level, and
then verifying the correctness of those properties for the code, all in an
automatic manner. While the nature of controllers and properties supported is
relatively restricted, this effort demonstrate the feasability of a paradigm
where domain specific knowledge is leveraged and automatically assists code
analysis. This opens the way for numerous directions of research. As the
mathematical breadth of theorem provers increase, more and more complex code
invariants can theoretically be handled, and thus more and more complex
controllers. One major area that should be explored is the integration of
the plant model at the level of the code, and the natural compromise that
arises between the precision the model and the finesse of the properties that
can be proven on the interconnection. This would pave the way for the
verification of a wide array of new properties, as most control theory results
are expressed on such interconnections between plant and controller. Soundness
of the results with respect to floating point computation is another issue that
requires attention.
We applied the toolchain to a mass-spring-damper system for its open-loop stability property.

\section*{Acknowledgements}
The authors would like to thank Pierre Roux for his contribution to the pvs-ellipsoid plugin, Gilberto Perez and Pablo Ascariz for their invaluable help on the PVS linear algebra library.

This article was prepared under support from the Army Research Office under MURI Award W911NF-11-1-0046, 
NSF Grant CNS - 1135955 “CPS: Medium: Collaborative Research: Credible Autocoding and Verification of Embedded Software (CrAVES)”, FUI 2011 project P,  the National Aeronautics and Space Administration under NASA Cooperative Agreement NNL09AA00A, activity
2736, ITEA2 OPES, FNRAE project CAVALE, ANR INS project CAFEIN and ANR ASTRID project VORACE.


\bibliographystyle{plain}
\bibliography{complete} 

\end{document}